\documentclass[reprint,prx,superscriptaddress,showpacs,twocolumn]{revtex4-1}
\usepackage{units}
\usepackage{amsmath}
\usepackage{amssymb}
\usepackage{graphicx}
\usepackage{bm}
\usepackage{microtype,color}

\newcommand{\be}{\begin{equation}}
\newcommand{\ee}{\end{equation}}
\newcommand{\bea}{\begin{eqnarray}}
\newcommand{\eea}{\end{eqnarray}}

\begin{document}

\title{Control of mesoscopic transport by modifying transmission channels in opaque media}

\author{Raktim Sarma}
\affiliation{Department of Applied Physics, Yale University, New Haven, CT, 06520, USA}
\author{Alexey Yamilov}
\email{yamilov@mst.edu}
\affiliation{\textls[-20]{Department of Physics, Missouri University of Science and Technology, Rolla, Missouri 65409, USA}}
\author{Seng Fatt Liew}
\affiliation{Department of Applied Physics, Yale University, New Haven, CT, 06520, USA}
\author{Mikhael Guy}
\affiliation{Science Research Software Core, Yale University, New Haven, CT, 06520, USA}
\author{Hui Cao}
\email{hui.cao@yale.edu}
\affiliation{Department of Applied Physics, Yale University, New Haven, CT, 06520, USA}

\date{\today}

\begin{abstract}

While controlling particle diffusion in a confined geometry is a popular approach taken by both natural and artificial systems, it has not been widely adopted for controlling light transport in random media, where wave interference effects play a critical role. The transmission eigenchannels determine not only light propagation through the disordered system but also the energy concentrated inside. Here we propose and demonstrate an effective approach to modify these channels, whose structures are considered to be universal in conventional diffusive waveguides. By adjusting the waveguide geometry, we are able to alter the spatial profiles of the transmission eigenchannels significantly and deterministically from the universal ones. In addition, evanescent channels may be converted to propagating channels by gradually increasing the waveguide cross-section. Our approach allows to control not only the transmitted and reflected light, but also the depth profile of energy density inside the scattering system. In particular geometries perfect reflection channels are created, and their large penetration depth into the turbid medium as well as the complete return of probe light to the input end would greatly benefit sensing and imaging applications. Absorption along with geometry can be further employed for tuning the decay length of energy flux inside the random system, which cannot be achieved in a common waveguide with uniform cross-section. Our approach relies solely on confined geometry and does not require any modification of intrinsic disorder, thus it is applicable to a variety of systems and also to other types of waves.

\end{abstract}


\maketitle

\section{Introduction\label{sec:intro}}

The diffusive transport of particles in a confined geometry can be effectively controlled by varying the boundary shape. This approach has been widely adopted in natural and artificial systems including channels in biological membranes, nanoporous materials, microfluidics, and artificial ion channels \cite{Zhou2008, Lindner2004, Hille2001, Beerdsen2006, Keil2000, Bruus, Hanggi2009}. A large variety of quasi one-dimensional (1D) structures with modulated cross-section have been developed for applications in controlled drug delivery, biochemical sensing, particle sorting, Brownian motors, and ion pumps \cite{Hanggi2009, Manz2006, Dekker2007, Zhao2012}. However, this powerful method has not been extensively applied to the control of $\textit{diffusive transport of waves}$ such as light, microwave or acoustic waves.

While wave diffusion is often described by the Brownian motion, it has a fundamental difference from particle diffusion, i.e., the scattered waves interfere and produce many important phenomena in mesoscopic physics, e.g., Anderson localization, universal conductance fluctuations, and enhanced backscattering \cite{Sheng2,Akkermanbook,VanRossum_RMP, 2009_Lagendijk_PT}. Our aim is to control the mesoscopic transport by manipulating wave interference effects in a confined geometry.

A prominent interference effect in a lossless diffusive medium is the creation of open and closed channels, which are eigenvectors of the matrix $t^{\dagger} t$, where $t$ is the field transmission matrix (TM). The transmission eigenvalues are close to 1 or 0, leading to a bimodal distribution \cite{Dorokhov1982, Dorokhov1984, Mello88, Nazarov94, 1986_Pichard_Eigenchannels, 1986_Imry, 1992_Pendry_math, Beenakker1, Stone1}. The open channels (with transmission eigenvalues $\tau$ close to 1) have dominant contributions to the propagation of waves through random media, while the closed channels ($\tau \sim 0$) determine the reflected waves. Thus by modifying these channels, one would be able to control wave transport. The key question is, then, how to modify these channels.

A recent study has shown that the maximum transmission channel has a universal spatial profile (inside a diffusive waveguide with uniform cross section), which cannot be changed by varying disorder strength or by adjusting the width or length of the random media \cite{Genack15}. The wavefront shaping technique has been successfully developed for selective coupling of light into open channels to enhance the total transmission or focusing through a random medium \cite{Vellekoop2008, Choi2011, Kim12, YuPRL13, Popoff2014}, but it cannot modify the transmission eigenchannels. Therefore, an efficient method for deterministic tailoring of the spatial structure of transmission channels is still missing.

In this paper, we propose and demonstrate an effective approach to manipulate the transmission eigenchannels to control diffusive wave transport. We show that by varying the geometry of a random waveguide, the spatial structure of open channels can be significantly and deterministically altered from the universal ones. This enables tuning the depth profile of energy density inside the random medium, thus controlling how much energy is concentrated inside the sample and where it is concentrated. By gradually increasing the waveguide cross-section, we are able to convert evanescent channels to propagating channels. In addition to controlling transmission, perfect reflection channels can be created in certain confined geometries, which do not exist in waveguides with uniform cross-section. We show that, unlike high reflection channels in uniform waveguides that exhibit shallow penetration into the disordered system, a perfect reflection channel can penetrate almost through the entire system but does not transmit any light. Furthermore, in the presence of absorption, we can vary the decay length of energy flux inside a diffusive waveguide by modulating the cross-section of the waveguide along its axis. This cannot be achieved  in a waveguide of uniform cross-section, as the flux decay length is independent of the waveguide dimension and is determined only by the intrinsic disorder and dissipation.

Optical absorption is ubiquitous and it often weakens the localization effects \cite{JohnPRL84, FreilikherPRL94, Brouwer, MisirpashaevPA97, ChabanovNat00, DeychPRB01, YamilovPRE04, Dz1, Tian1}, but our approach of using geometry to control wave transport by manipulating the structure of eigenchannels proves to be effective and robust against strong absorption. Therefore the confined geometries enable us to control not only the amount of light being transmitted or reflected, but also the amount of energy concentrated inside the random media. Although strong localization effects, absorption or asymmetric reflection from edges can modify the universal structure of transmission channels, but such effects also remove the open channels with perfect transmission \cite{Brouwer, Tianeigenchannel, Liew2014}. Unlike these effects, the approach of varying shape of confined geometries gives the significant advantage and freedom to alter the spatial structures of eigenchannels while retaining the open eigenchannels with perfect transmission.

Aside from the fundamental importance, the ability of tailoring the spatial distribution of energy density of transmission eigenchannels can be exploited to manipulate light-matter interactions in highly scattering media, e.g., light absorption, emission, amplification, and nonlinear optical processes \cite{Genack15, Cheng14}. The potential applications range from laser surgery, photovoltaics, to random laser and energy-efficient lighting \cite{Gratzel03, Vellekoop2010, Popoff_2010, MoskNatPhoReview, CaoJPA05, Vos13, Liew_solarcell, KimOE15}. Our results suggest that the perfect reflection channels may greatly benefit sensing and imaging applications, as the light in such a channel would penetrate to a certain depth and then fully reflected to ensure an efficient collection of the probe signal. The conversion of evanescent waves to propagative waves and vice versa may be used to tailor optical excitations inside the random media. Since the application of wavefront shaping technique to focusing or imaging through turbid media as well as enhancing total transmission depends on the properties of high transmission channels, our approach of modifying the transmission eigenvalues and eigenvectors by geometry provides a complementary degree of control. While the efficiency of wavefront shaping approach is reduced by incomplete channel control and measurement noise \cite{Popoff2014, Hasan13, Arthur1}, our approach is immune to such external factors. Although the above results are obtained for light, they are also applicable to other classical and quantum mechanical waves.

\section{Quasi-two-dimensional random waveguide}

To manipulate transmission eigenchannels, we design and fabricate quasi two-dimensional (2D) waveguides of various geometries. The waveguide structures are fabricated in a 220 nm silicon layer on top of 3 $\mu$m buried oxide by electron beam lithography and reactive ion etching \cite{Dz}. Figure 1 shows the scanning electron microscope (SEM) images of two fabricated waveguides. The waveguide contains a 2D random array of air holes that serve as scatterers for light. The air hole diameter is 100 nm and the average (center-to-center) distance of adjacent holes is 390 nm. The waveguide walls are made of triangle lattice of air holes (lattice constant = 440 nm, hole radius = 154 nm) that has a complete 2D photonic bandgap for the in-plane confinement of light. The waveguide is connected to a lead which is an empty waveguide (without any air holes) with a constant width to couple light in.

\begin{figure}[htbp]
\centering
\includegraphics[width=3.5in]{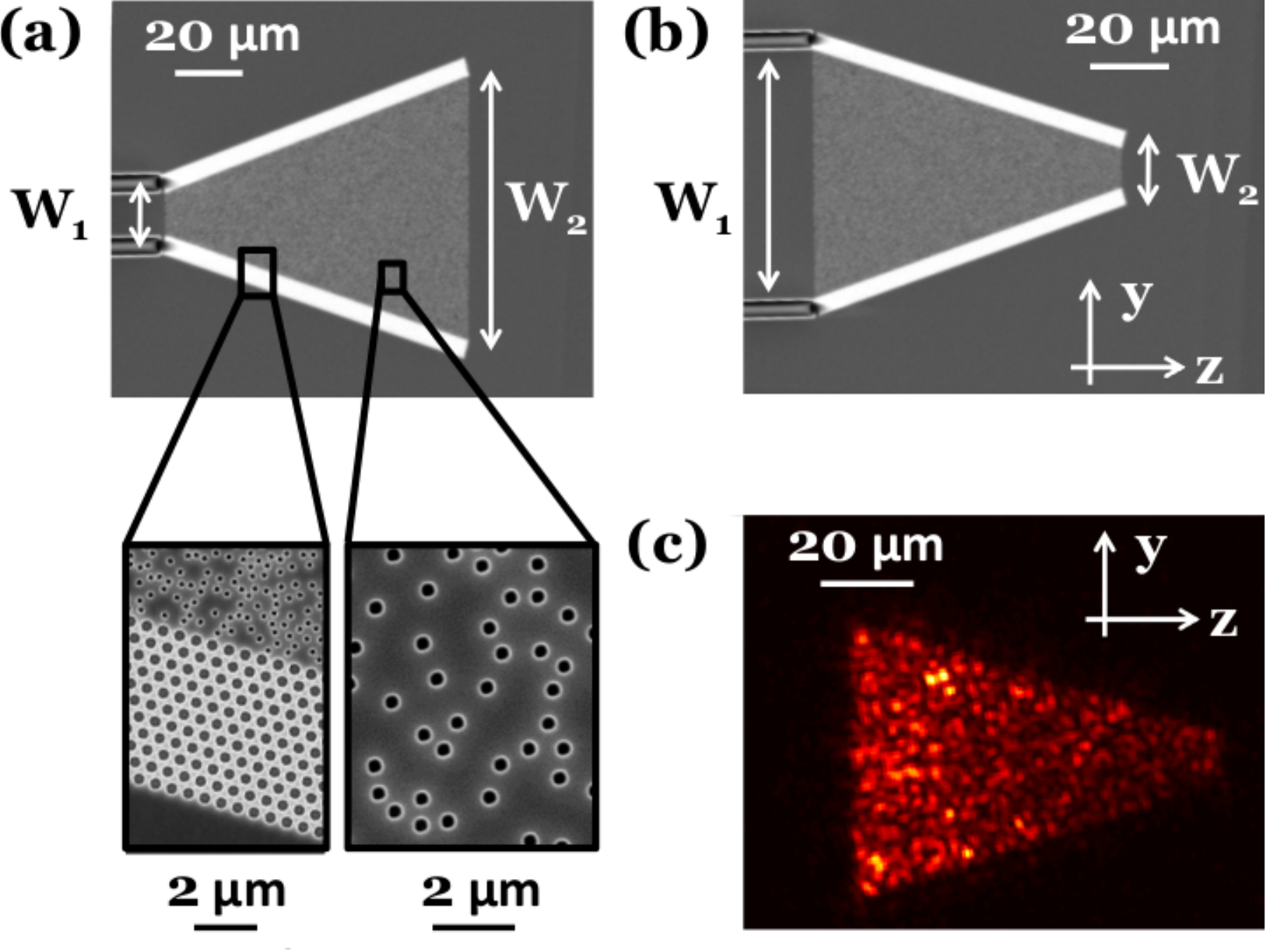}
\caption{Quasi-two-dimensional random waveguides of different geometry. (a,b) Top-view SEM images of fabricated quasi-2D disordered waveguides with linearly increasing (a) or decreasing (b) width. The width of waveguide in (a) increases from $W_1=10$ $\mu$m to $W_2=60$ $\mu$m, and in (b) it is opposite. Both have the same length $L$ = 80 $\mu$m. Magnified SEM images show the air holes distributed randomly in the tapered section of the waveguide and the triangle lattice of air holes in the reflecting sidewalls. (c) An optical image of the intensity of scattered light from the disordered waveguide. The wavelength of the probe light is 1500 nm. }
\end{figure}

A monochromatic beam of light from a tunable CW laser source (HP 8168F) is focused by an objective lens (numerical aperture NA = 0.4) onto the lead waveguide. The light is transverse-electric (TE) polarized, with the electric field parallel to the plane of the waveguide ($y-z$ plane). After propagating through the lead, the light is incident onto the random array of air holes and undergoes multiple scattering in the 2D plane of waveguide. Some of the light is scattered out of plane, part of which is collected by a 50$\times$ objective lens (NA = 0.42) and imaged onto an InGaAs camera (Xeva 1.7-320). From the optical image [Fig. 1(c)], the spatial distribution of light intensity inside the waveguide $I(y,z)$ is extracted. Ensemble averaging is done by recording the intensity profile for 50 different wavelengths around $\lambda$ = 1500 nm and three distinct configurations of air holes. Further averaging is done by slightly shifting the input beam spot on the lead waveguide in the transverse direction $y$ to produce distinct speckle illumination for the random array of air holes, nevertheless, the incident intensity profile is always kept uniform across $y$.

All disordered waveguides studied in this work exhibit diffusive transport. The relevant parameters for light transport in the disordered waveguide are the transport mean free path $\ell$ and the diffusive dissipation length $\xi_a$. The transport mean free path $\ell$ depends on the density and diameter of the air holes. The dissipation results from out-of-plane scattering, since the silicon absorption at the probe wavelength is negligible. This vertical leakage of light can be treated similarly as absorption and described by the diffusive dissipation length $\xi_a = \sqrt{D \tau_a}$, where $\tau_a $ is the ballistic dissipation time and $D$ is the diffusion coefficient \cite{Dz}. The values of $\ell$ and $\xi_a$ are 2.2 $\mu$m and 26 $\mu$m respectively, which were extracted from the measured intensity distribution inside a waveguide of rectangle shape \cite{Dz}. Since these two parameters depend only on the size and density of the air holes, we keep them the same for all waveguides with different geometries. This ensures the modification of light transport is purely due to the change in geometry instead of structural disorder or dissipation.

\section{Linear tapering of waveguide width \label{sec:simulations1}}

In Fig. 1, the two waveguides have their width $W(z)$ increase or decrease linearly along the waveguide axis $z$. To illustrate how the transmission channels are modified by the linear tapering of the waveguide boundary, we first perform numerical modeling by excluding the effect of dissipation. This enables us to separate the effect of geometry from that of dissipation, which  will be discussed in the next section. In the simulation, the wavelength, refractive index, and polarization of light are the same as in the experiment. However, the dimension of the waveguide and the transport mean free path are scaled down to reduce the computing time. This should not change the conclusion of our results because the systems are still in the diffusive regime.

The disordered waveguide has perfectly reflecting sidewalls and is connected to two leads (empty waveguides) at both ends. The refractive index in the empty waveguide is determined by the vertical waveguiding in the silicon layer, and its value is calculated to be $n$ = 2.85. In the disordered waveguide, the presence of air holes ($n =1$, radius = 75 nm, filling fraction = 0.15) reduces the effective index of refraction to  $n$ = 2.62. The (vacuum) wavelength of the probe light is $\lambda$ =1.5 $\mu$m, and the transport mean free path is $\ell$ = 1.1 $\mu$m. The length of the disordered waveguide $L$ is set to be much larger than $\ell$ to ensure multiple scattering and diffusion of light. Since the localization length ($\xi$) is proportional to the width of the waveguide ($W$), the value of $W$ is chosen to make $\xi \gg L$ so that localization effects are negligible.

We calculate the electromagnetic field inside the disordered waveguide by solving the Maxwell's equations using the finite difference frequency-domain method (COMSOL Multiphysics). To construct the transmission matrix $t$ of the disordered waveguide, we use the guided modes in the leads as the basis. The input (output) lead waveguide has a constant width equal to the same width $W_1$ ($W_2$) of the disordered waveguide at the front (back) end $z=0$ ($z=L$), and it supports $M = W_1 /\lambda/2n$ ($N= W_2 /\lambda/2n$) guided modes. Thus $t$ is a $N \times M$ matrix, and its element $t_{ij}$ represents the field transmission from the input $j$-th mode to the output $i$-th mode. The reflection matrix is constructed in a similar way by computing the reflected waves, and its dimension is $M\times M$.

A singular value decomposition of the transmission matrix $t$ gives $t = U \Lambda V^{\dagger}$. $\Lambda$ is a $N \times M$ diagonal matrix with $min[N,M]$ non-negative real numbers, $\sqrt{\tau_m}$, where $\tau_m$ is the eigenvalue of $t^{\dagger}t$ and represents the transmittance of the $m^{th}$ transmission eigenchannel. $V$ is a $M \times M$ unitary matrix that maps the field in the guided modes of the input lead to the eigenchannels of the disordered waveguide, and $U$ is a $N \times N$ unitary matrix that maps the eigenchannels to the output waveguide modes. Each column of $V$ represents an input singular vector, whose elements are the complex coefficients for the input waveguide modes that combine to couple light into a single transmission eigenchannel. The output field of a single transmission eigenchanel is represented by the column of $U$, which is called the output singular vector. Similarly the reflection eigenvalues $\rho_n$ can be obtained by singular value decomposition of the reflection matrix $r$.

For comparison, we also compute the transmission eigenchannels in the waveguide of constant width $W$. For $W$ = 5.1 $\mu$m, $M$ = $N$ = 19, and there are 19 transmission eigenchannels. Figure 2(a) plots the transmission eigenvalues, 2 of which ($m$ = 18 and 19) are many orders of magnitude smaller than the others and are not shown as they fall below the numerical accuracy . This is because the lead waveguide has larger refractive index than the disordered waveguide and support more guided modes. The disordered waveguide can support only $N-2$ = 17 propagating modes, thus 2 of the 19 transmission channels cannot propagate inside the disordered waveguide and become evanescent. Light can be coupled to these two evanescent channels with the extra modes that can be supported by the input lead waveguide.

Therefore, the eigenchannels of the transmission matrix can be divided into two categories: propagating channels and evanescent channels. The propagating channel has a spatial structure that varies on the scale of the mean free path. The evanescent channel features an intensity decay on the order of the wavelength, which is much shorter than the mean free path, and the corresponding transmission eigenvalue is essentially zero.

A gradual increase of the waveguide width along its axis increases the number of propagating modes that can be supported inside the disordered waveguide, converting the evanescent channels to the propagating channels. This is observed, as an example, in the tapered waveguide whose width is increased from $W_1$ = 5.1 $\mu$m at $z=0$ to $W_2$ = 10.2 $\mu$m at $z=L$ [Fig. 1(a)]. With $M= 19$ and $N =38$, the transmission matrix $t_{38 \times 19}$ still supports $19$ transmission eigenchannels, but all of them have non-vanishing $\tau_m$ [Fig. 2(a)].

Figure 2(b) shows the spatial distribution of electric field intensity inside the tapered waveguide for the 18th and 19th transmission eigenchannels which have the lowest transmittance. Both these channels have been converted from evanescent channels in a constant-width waveguide with $W$= 5.1 $\mu$m to propagating channels in the tapered waveguide. $I(y,z)$ exhibits a sharp drop near the front side of the waveguide. For a quantitative analysis, the cross-section-averaged intensity, $I_v(z) = [1/W(z)] \int I(y,z) dy$, is plotted in Fig. 2(c) for these two channels. For comparison, $I_v(z)$ for the $N=10$ eigenchannel is added to the plot, and it displays an exponential decay with a constant rate. In contrast, $I_v(z)$ for the 19th eigenchannel first decays very rapidly at small $z/L$, and then changes to a much slower decay at $z/L \sim$ 0.07. The number of guided modes in the waveguide is $N(z) = 2W(z)/(\lambda/n)$, where $W(z)$ is the waveguide width at depth $z$,  and $n$ is the effective index of refraction of the disordered waveguide. As $W$ increases with $z$, the waveguide becomes wide enough to support additional modes. For example, at $z/L \sim$ 0.07, $N$ is increased from 18 to 19, thus the 19th mode is transformed from evanescent wave to propagating wave. Consequently, the decay length of $I_v(z)$ increases from $\sim$ 0.14 $\mu$m (comparable to $\lambda/2 \pi n$) to $\sim$ 1.8 $\mu$m (much larger than $\lambda/2 \pi n$). Similarly, the 18th eigenchannel is transformed from evanescent to propagating at a smaller value of $z/L \sim$ 0.05, where $N$ is increased from 17 to 18. Hence, this conversion can be attributed to the gradual increase of the number of propagating modes that can be supported by the tapered waveguide at different depths.

If $\lambda$ and $n$ are fixed, the spatial position ($z/L$) inside the tapered waveguide where the conversion from evanescent wave to propagating wave takes place is determined by the width at that position, thus the spatial position where such conversion occurs can be easily controlled by tuning the tapering angle. The disorder strength does not affect directly the location of conversion, however, a change in the disorder strength is often accompanied by a change in the effective index of refraction $n$, which would modify the conversion depth.

\begin{figure}[htbp]
\centering
\includegraphics[width=3.5in]{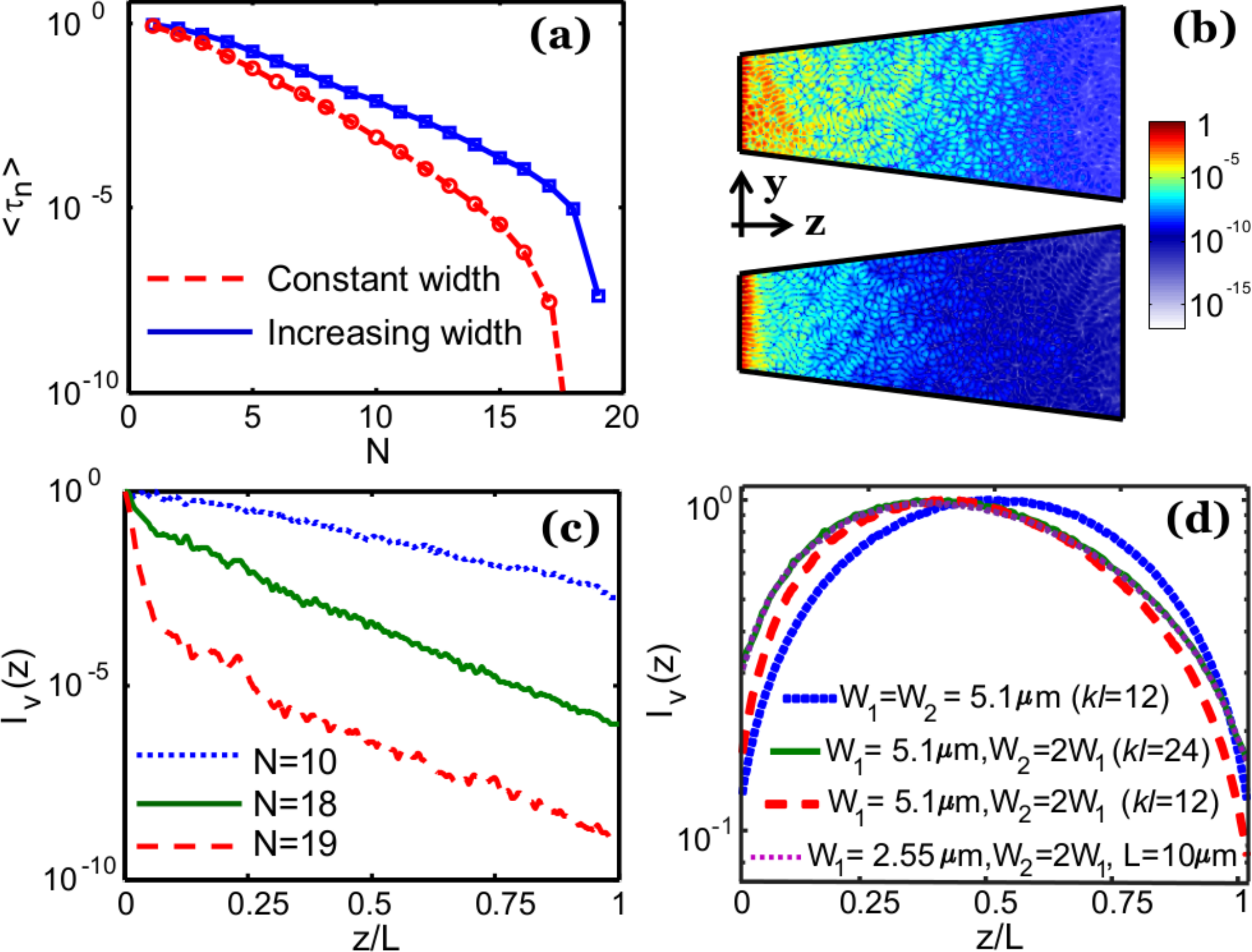}
\caption{(Color online) Comparison of transmission eigenvalues and eigenchannels in constant-width and increasing-width random waveguides. (a) Numerically-calculated ensemble-averaged transmission eigenvalues of random waveguides with constant width (dashed line with circles) and increasing width (solid line with squares). The constant-width waveguide ($W=5.1$ $\mu$m, $L=20$ $\mu$m) supports 19 transmission eigenchannels of which 17 are propagating channels and 2 are evanescent channels, whereas the expanding waveguide ($W_1=5.1$ $\mu$m, $W_2=10.2$ $\mu$m, $L=20$ $\mu$m) has 19 propagating channels of higher transmittance. (b) Spatial distribution of electric field intensity inside the waveguide with increasing width for the 18th and 19th transmission eigenchannels. Both transform from evanescent waves at the entrance of the waveguide to propagating waves due to the increase of waveguide width. (c) Cross-section-averaged intensity, $I_v(z)$, for the 18th (solid line) and 19th (dashed line) channels shown in (b). The conversion from evanescent wave to propagating wave causes a sudden change in the decay length of $I_v(z)$ near the front end of the waveguide. For comparison,  $I_v(z)$ for the 10th eigenchannel (dotted line) of the same waveguide is added and it shows a constant decay length. (d) Comparison of the cross-section-averaged intensity, $I_v(z)$, of the maximum transmission channel ($m = 1$) in the disordered waveguides with constant width (blue dotted line) and increasing width with two different disorder strengths (red dashed line and green solid line) and different dimensions (dotted magenta line). Tapering of the waveguide width breaks the symmetry of the spatial structure of the open channel and moves the peak of  $I_v(z)$ from the center of the waveguide towards the front end. The position of the peak does not depend on the disorder strength.
}
\end{figure}

The increase of the waveguide width also enhances the transmittance of all other transmission eigenchannels (albeit not as large an enhancement as the above two) which are also propagating channels in the constant width waveguide. Consequently, the dimensionless conductance $g = \sum \tau_m$ is larger, but the number of input modes remains the same. This behavior is distinct from the constant-width waveguide, where the increase of width also enhances $g$, but the number of input modes increases simultaneously requiring additional degree of control of the input field for coupling into a single eigenchannel. The waveguide with increasing width can therefore be useful for applications related to enhancing transmission through random media by wavefront shaping technique with incomplete degree of control of the input field.

Furthermore, the spatial profiles of open channels are modified in the tapered waveguide. Figure 2(d) compares the cross-section-averaged intensity $I_v(z)$ of the maximum transmission channel in the disordered waveguides with constant and increasing widths. In the waveguide with uniform cross-section, $I_v(z)$ exhibits a symmetric profile with peak in the middle of the waveguide ($z=L/2$). It corresponds to the universal structure of the maximum transmission channel in a constant-width waveguide \cite{Genack15}. In the waveguide with increasing widths,  $I_v(z)$  becomes asymmetric and its peak shifts from the center towards the front end of the waveguide ($z<L/2$). As seen in Fig. 2(d), when the tapering angle of the waveguide boundary is merely $14^{\circ}$, the peak of the maximum transmission channel has already moved significantly from the center $z/L = 0.5$ to $z/L = 0.35$.  This shift does not depend on the disorder strength or the actual dimension of the diffusive waveguide. As a confirmation, Fig. 2(d) shows the spatial profiles of the highest transmission channel in two more tapered waveguides, one has $L$, $W_1$, $W_2$ all reduced to half, but $k \ell$ unchanged ($k = 2 \pi / \lambda / n$); the other has the same $L$, $W_1$, $W_2$, but $k \ell$ is doubled.  Although their profiles are slightly different, the peak positions are identical.

Next we investigate the disordered waveguide with linearly decreasing width, as shown in Fig. 1(b). This geometry is the mirror image of the one in Fig. 1(a), thus light injection from the left end of waveguide in Fig. 1(b) is identical to light injection from the right end of the waveguide in Fig. 1(a). The transmission matrix of the waveguide in Fig. 1(b), $t_{19 \times 38}$, is the transpose of that in Fig. 1(a), and it also supports 19 transmission eigenchannels with the same transmittance. Thus the conductance $g$ is identical for the two waveguides in Fig. 1. However, the spatial structure of the open channels is different.

Figure 3(a) shows the cross-section-averaged intensity $I_v(z)$ for the maximum transmission channel in the waveguide with decreasing width. Its peak shifts from the center of the waveguide towards the output end ($z>L/2$), opposite to that of the waveguide with increasing width in Fig. 2(d). The two profiles are mirror image, and the peak always shifts towards the narrower section of the tapered waveguide. How much the peak shifts from the waveguide center depends on the angle of tapering. By changing the tapering angle, the location of the intensity peak can be tuned deterministically, as seen in Fig. 3(a). This result illustrates that the maximum of the energy density can be positioned to different depths inside a random system by tailoring its geometry.

While the number of the transmission eigenchannels for the two waveguides in Fig. 1 is identical, the number of reflection channels differs.
In the expanding waveguide [Fig. 1(a)], the reflection matrix $r_{19 \times 19}$ has 19 eigenchannels, which have one-to-one correspondence with the transmission eigenchannels. However, in the contracting waveguide [Fig. 1(b)], the input lead waveguide supports 38 guided modes, and the output only 19 modes. Consequently there are 19 transmission eigenchannels, but 38 reflection channels. While 19 of the reflection channels have the corresponding transmission channels, the rest 19 do not. In other words, the reflection matrix $r_{38 \times 38}$ has 38 eigenvalues, of which 19 of them are equal to unity. They represent perfect reflection channels with all incident light being reflected.

The 2-D spatial distribution of field intensity for a perfection reflection channel in the tapered waveguide with $W_1/W_2=2$ is shown in the inset of Fig. 3(b). For comparison, a high reflection channel in a waveguide of uniform cross-section is also shown. We can clearly see that the high reflection channel in the constant-width waveguide has a uniform decay of intensity inside the random structure. In contrast, the intensity of the perfect reflection channel exhibits a much slower decay almost throughout the entire random structure and then a sharp drop close to the rear end ($z\sim L$).

The main panel of Fig. 3(b) plots the cross-section-averaged intensity, $I_v(z)$, for one of the perfect reflection channels in two tapered waveguides with different tapering angles and a high reflection channel in the constant-width waveguide. The high reflection channel of a constant-width waveguide has shallow penetration into the random medium due to a rapid intensity decay. The perfect reflection channel, however, has a much slower decay and thus a longer penetration depth. A sharp drop of its intensity near the rear end corresponds to the cutoff beyond which no light propagates. The cutoff occurs at the position where the waveguide width is just large enough to support $N+1$ modes (where $N$ is the number of propagating modes in the output lead). Since the cutoff position depends on the tapering angle of the random waveguide, both the decay length of the intensity and the cutoff position in a perfect reflection channel can be deterministically and effectively controlled by tuning the tapering angle. For example, by increasing the tapering angle we are able to increase the penetration depth by shifting the cutoff position closer to the output end, as seen in Fig. 3(b).

Since light in the perfect reflection channels can penetrate deep into the scattering system, such channels can be used for probing inside turbid media. Despite of the deeper penetration, all the light exits from the input end, making the collection efficiency of probe signal $100\%$, which is extremely useful for sensing or imaging applications. The  penetration depth can be precisely tuned via tapering the boundary of a confined random system.

\begin{figure}[htbp]
\centering
\includegraphics[width=3.5in]{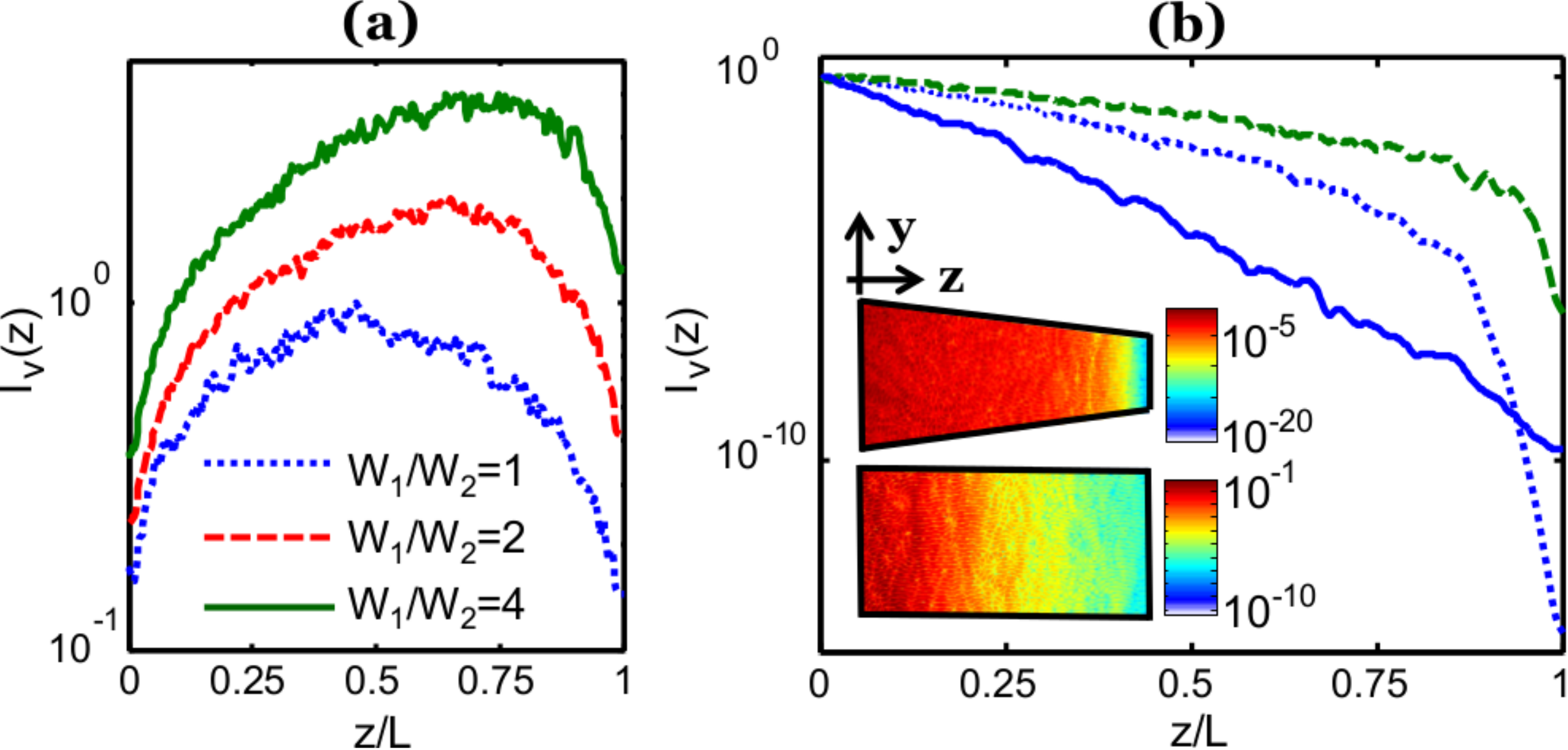}
\caption{(Color online) Transmission eigenchannels in tapered waveguide of decreasing width. (a) Comparison of cross-section-averaged intensity, $I_v(z)$, of the maximum transmission channel ($m=1$) in two waveguides with different tapering angles and a constant-width waveguide. All waveguides have the same length $L=20$ $\mu$m. The constant-width waveguide has $W_1=W_2=10.2$ $\mu$m (blue dotted line). The two tapered waveguides have $W_1=10.2$ $\mu$m and $W_1/W_2=2$ (red dashed line),  $4$ (green solid line). The $I_v(z)$ curves are offset along the y axis for clarity. The intensity peak shifts from the waveguide center ($W_1/W_2=1$)  towards the output end ($W_1/W_2>1$), and the shift is larger for higher tapering angle (larger $W_1/W_2$). (b) Cross-section-averaged intensity, $I_v(z)$, of a perfect reflection channel for the same tapered waveguides as in (a). $I_v(z)$ of a high reflection channel of the constant-width waveguide (blue solid line) is added for comparison. The insets show the spatial distribution of electric field intensity for the high reflection channel of the constant-width waveguide and the perfect reflection channel of the tapered waveguide with $W_1/W_2=2$. The perfect reflection channel in a tapered waveguide exhibits slower intensity decay inside the random medium (followed by a sharp drop near the rear end) and thus can penetrate much deeper into the turbid medium than the high reflection channel in the constant-width waveguide. The penetration length increases with the tapering angle.
}
\end{figure}

\section{Effect of Absorption \label{sec:simulations2}}

In this section, we study the effect of light dissipation, which was not included the last section. Previous studies have shown that loss has a profound impact on the transmission channels. It not only modifies the statistical distribution of the transmission eigenvalues \cite{Brouwer}, but also changes the structure of eigenchannels \cite{Liew2014, Liew2015}. In case of passive diffusive waveguides, the probability density distribution of transmission eigenvalues, $P(\tau)$, has 2 peaks, one at 1 and the other at 0. Thus, there are many channels (the number is determined by $g$) with comparable values of $\tau \sim ~1$. Therefore, the total intensity inside the random medium is the sum of intensities of all these high transmission eigenchannels.

In case of strongly absorbing waveguides ($L\gg\xi_a$), the peak at $\tau =1$ disappears and $P(\tau)$ has a cutoff at $\tau_{max}$ which is determined by $L/\xi_a$. In such absorbing waveguides, $P(\tau)$ decays strongly with $\tau$ with a faster decay near $\tau_{max}$. This implies that the $\tau$'s will be arranged as $\tau_1>\tau_2>\tau_3>..$. Furthermore, because $P(\tau)$ decays fast toward $\tau_{max}$, $\tau_1$ will be much greater than $\tau_2, \tau_3, \tau_4...$ and the total intensity inside the random media will therefore be dominated by the eigenchannel with the maximum transmission.

For the quasi-2D waveguides we fabricate, the dissipation results from out-of-plane scattering of light, which can be treated as absorption \cite{Dz}.
We simulate it in the 2D waveguide by introducing an imaginary part of the refractive index. The diffusive dissipation length is $\xi_a = \sqrt{\ell l_a/2}$,  where $l_a$ is the ballistic dissipation length. The ratio of $L/\xi_a$ is set to 3.0, which is close to the value of the fabricated waveguides. At $L/\xi_a = 3$, absorption causes a notable change in the spatial profile of the maximum transmission channel as seen in Figure 4(a). The intensity peak of the maximum transmission channel, which is located at the middle ($z/L \sim 0.5$) of the passive waveguide, moves to the front end ($z/L \sim 0$) due to absorption.

Although it reduces the throughput, loss allows us to manipulate the spatial decay of energy flux inside the random waveguide by geometry.
In the absence of loss, the net flux $J(z)$, integrated over the cross-section of the waveguide, points in the $z$ direction and its value is constant along $z$. By tailoring the boundary shape of the waveguide, the magnitude of $J$ changes, but it remains invariant with $z$.
With the addition of loss, $J(z)$ decays exponentially along $z$. If the waveguide has a uniform cross-section, the decay length is determined by $\xi_a$,  which is independent of the waveguide width or length. However, the decay length can be varied by tapering the waveguide width along $z$. Figure 4(b) plots $J(z)$ in four waveguides with random input fields. To compare the spatial profile of $J(z)$, its value at $z=0$ is normalized to 1. Two of the waveguides have uniform width, $W$ = 5.1 $\mu$m, 10.2 $\mu$m, and their $J(z)$ overlaps after the normalization. With a linear increase of $W$ with $z$, the decay of $J(z)$ becomes slower, while a linear decrease of the waveguide width accelerates the flux decay. Hence, by varying the waveguide width along the cross-section, we can tune the decay of energy flux inside the random media. Such tuning of flux decay rate by geometry can be achieved only in the presence of loss, illustrating additional degree of control enabled by combination of dissipation and geometry.

\begin{figure}[htbp]
\centering
\includegraphics[width=3.5in]{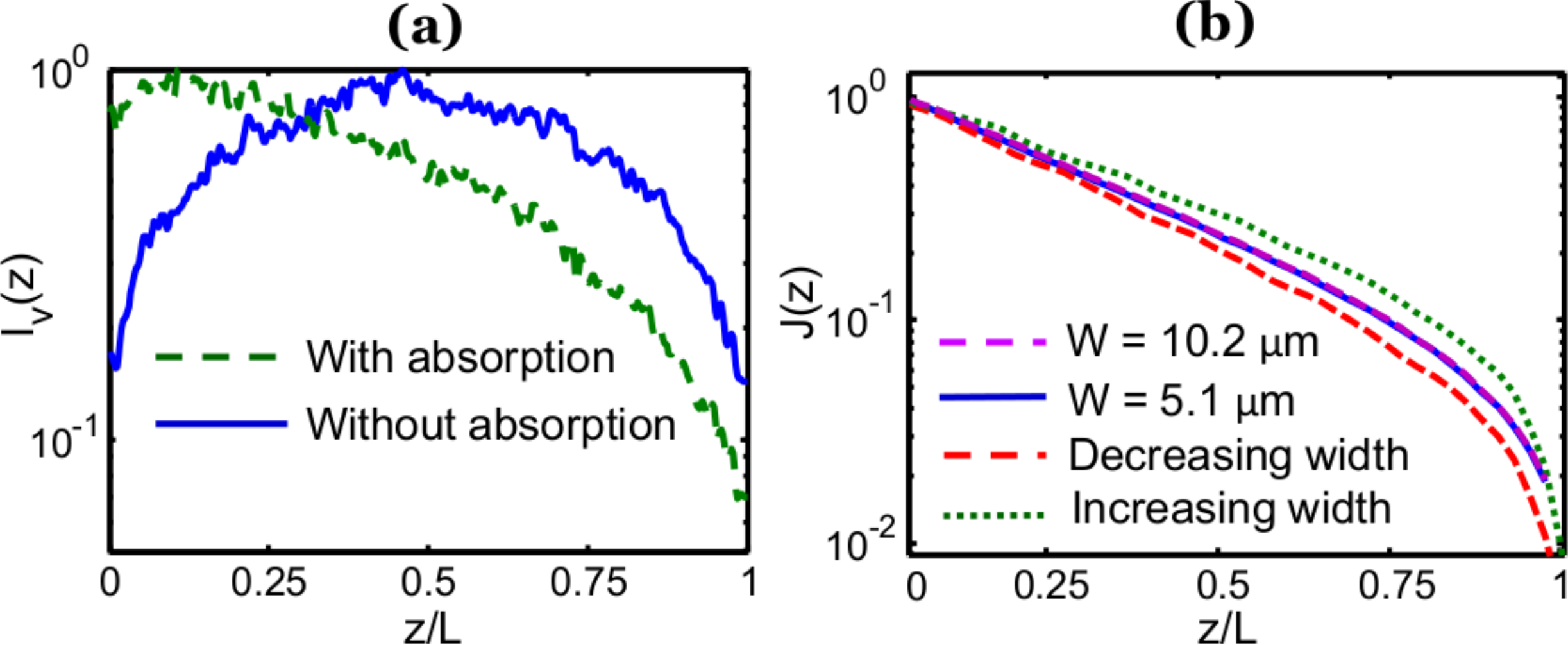}
\caption{(Color online) Effect of absorption on the maximum transmission eigenchannel and energy flux decay in the constant-width and tapered waveguides. (a) Comparison of the cross-section-averaged intensity, $I_v(z)$, of the maximum transmission channel in a constant-width ($W=5.1$ $\mu$m, $L=20$ $\mu$m) disordered waveguide with (dashed line) and without (solid line) absorption. In the absorbing waveguide, $L/\xi_a=3$. Absorption modifies the spatial profile of the maximum transmission channel. (b) Comparison of spatial decay of energy flux $J(z)$ in random waveguides with constant widths $W=5.1$ $\mu$m (blue solid line) and $W=10.2$ $\mu$m (dashed magenta line), increasing width of $W_1=5.1$ $\mu$m, $W_2=10.2$ $\mu$m (green dotted line) and decreasing width of $W_1=10.2$ $\mu$m and $W_2=5.1$ $\mu$m (red dashed line). For all waveguides, $L=20$ $\mu$m, $L/\xi_a=3$ and $J(z)$ is normalized to 1 at $z=0$. While the flux decay length remains the same for the two rectangle waveguides of different widths, it is lengthened in the expanding waveguide and shortened in the contracting waveguide.
}
\end{figure}

\section{Intensity decay inside random media \label{sec:experiment}}

Experimentally we measured the 2D intensity distribution inside the tapered waveguides shown in Fig. 1(c). From $I(y,z)$ we obtain the cross-section-averaged intensity $I_v(z)$ and the cross-section-integrated intensity $I_t(z) = I_v(z) \, W(z)$. The former gives the depth profile of the average energy density inside the random waveguide, and the latter tells the total amount of energy concentrated at certain depth $z$. In the tapered waveguides, the intensity decay rates become significantly different as seen in Fig. 5(a,b). $I_t(z)$ (total amount of energy concentrated at certain depth $z$) decays much slower inside the expanding waveguide than that in the constant-width waveguide, while the contracting waveguide leads to a much faster decay of $I_t(z)$. $I_v(z)$ (depth profile of the average energy density), however, displays an opposite behavior: it decays faster in the expanding waveguide and slower in the contracting waveguide. Such behavior is attributed to the variation of the waveguide width along $z$.

For comparison, we also measure the intensity decay inside two constant-width waveguides. Despite of a factor of 6 difference in the waveguide width ($W$=10 $\mu$m, 60 $\mu$m), $I_v$ and $I_t$ decay exponentially in the two waveguides with nearly the same rate (not shown). This result confirms that the intensity decay is independent of the waveguide width  as long as $W$ is invariant with $z$ and localization effect is negligible \cite{Dz}.

As mentioned before, the two tapered waveguides with the same tapering angle are mirror image of each other with respect to $z=L/2$. Thus the transport of light with input from one end ($z=0$) of one waveguide is equivalent to that with input from the opposite end ($z=L$) of the other waveguide. Hence, the difference in the intensity decay in the two waveguides with injection from the same end ($z=0$) illustrates asymmetric transport of light in such tapered waveguides.

Since our fabricated waveguides are in the regime of strong dissipation ($L\gg\xi_a$), the intensities inside the structures are dominated by the maximum transmission channel. The experimentally measured intensities should therefore reflect qualitatively the intensity profiles of the maximum transmission channels. In Figure 5(c,d), we plot the numerically calculated $I_v(z)$ and $I_t(z)$ for the maximum transmission eigenchannels in waveguides of constant widths and tapered geometries (with reduced dimensions due to limited computing power). They exhibit qualitatively similar structures, indicating the intensity distribution inside a strongly dissipative random system is determined by the maximum transmission channel whose spatial profile can be tuned by geometry.

\begin{figure}[htbp]
\centering
\includegraphics[width=3.5in]{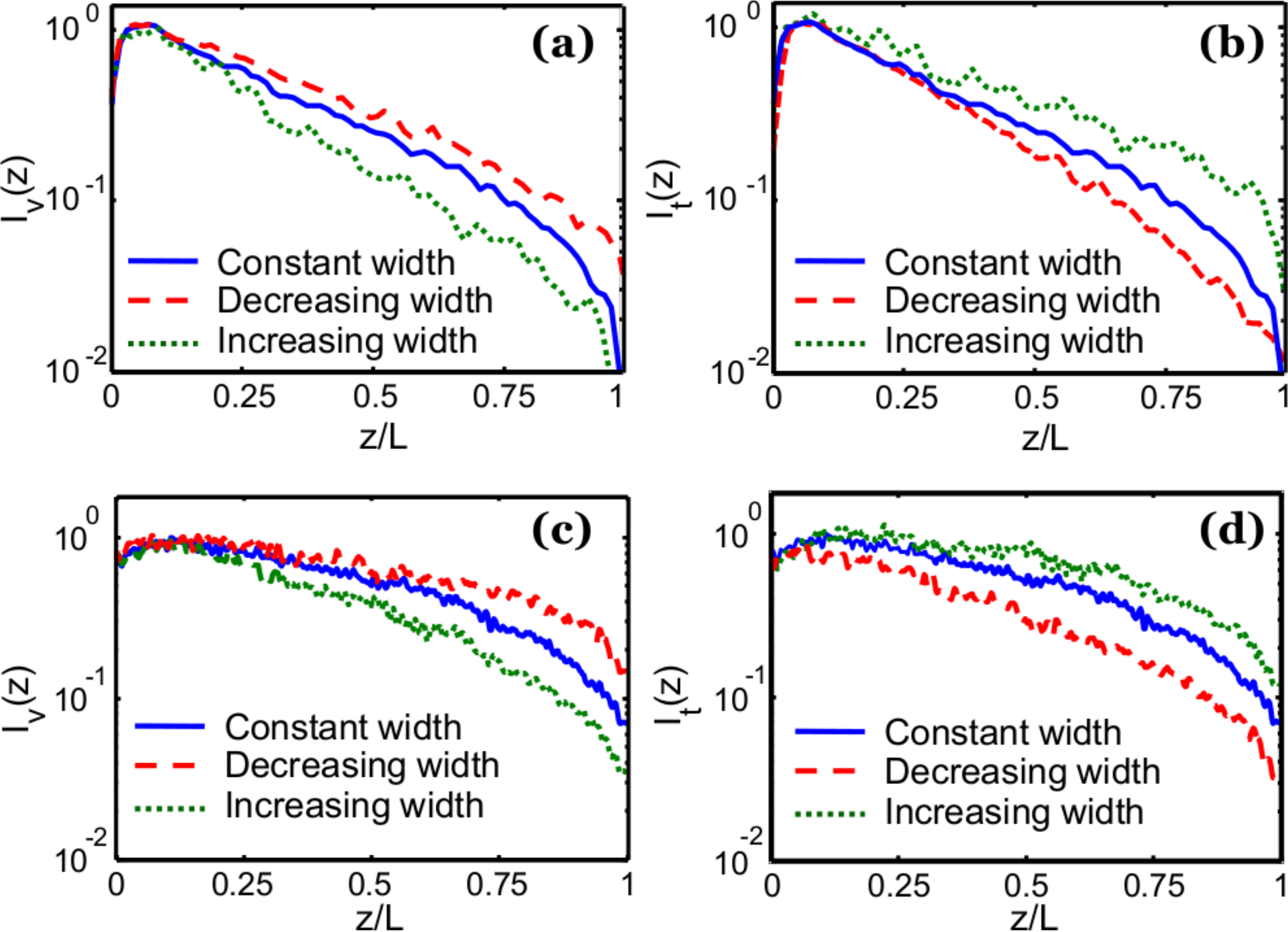}
\caption{(Color online) Experimentally measured intensity decays inside disordered waveguides in comparison to numerically calculated spatial profiles of the maximum transmission eigenchannels. (a, b) Experimentally measured cross-section-averaged intensity $I_v(z)$ (a) and cross-section-integrated intensity $I_t(z)$ (b) inside quasi-2D waveguides of constant width $W=60$ $\mu$m  (solid blue line), increasing width with $W_1=10$ $\mu$m, $W_2=60$ $\mu$m (dotted green line), and decreasing width with $W_1=60$ $\mu$m,  $W_2=10$ $\mu$m (dashed red line). All the waveguides have $L=80$ $\mu$m and $L/\xi_a=3$. The tapering of the waveguide boundary causes an opposite change in the decay length of $I_v(z)$ and  $I_t(z)$. (c, d) Numerically calculated $I_v(z)$ (c) and $I_t(z)$ (d) of the maximum transmission eigenchannel in the disordered waveguides of constant width $W=10.2$ $\mu$m  (solid blue line), increasing width with $W_1=5.1$ $\mu$m, $W_2=10.2$ $\mu$m (dotted green line), and decreasing width with $W_1=10.2$ $\mu$m, $W_2=5.1$ $\mu$m (dashed red line). All waveguides have $L=20$ $\mu$m and $L/\xi_a=3$. Despite of the reduced waveguide dimensions, the maximum transmission channels exhibit a qualitatively similar structure to the experimentally measured intensities, indicating the intensity distribution inside a strongly absorbing random medium is determined by the structure of the maximum transmission channel.
}
\end{figure}

\section{Non-monotonic variation of waveguide cross-section \label{sec:bowtie}}

Finally we change the waveguide width non-monotonically along the axis for further control of transmission channels. Figure 6(a) shows a ``bow-tie'' waveguide whose width $W$ decreases linearly in the first half and then increases in the second half. While the input and output ends have identical widths, the waveguide has a constriction in the middle that reduces the energy flow. The total number of transmission eigenchannels is still determined by the waveguide width at the input/output. However, only a fraction of these channels (determined by the width of the constriction) can propagate through the constriction. The rest are converted to evanescent waves in the vicinity of the construction due to the reduction in the number of propagating modes. As the waveguide width increases after the constriction, the evanescent wave that can tunnel through the constriction may convert back to propagating wave. This is seen in the intensity profiles of the transmission channels in Fig. 6(b). $I_v(z)$ decays gradually in the first part of the bow-tie waveguide, then suddenly changes to a must faster decay near the constriction, after the constriction the decay slows down again. The abrupt changes in the decay length, from much larger than the evanescent decay length, $\lambda/2 \pi n$, to smaller than $\lambda/2 \pi n$ and back, indicate the conversion from propagating wave to evanescent wave and back. The accelerated decay rate near the constriction differs from one channel to another [Fig. 6(b)]. Hence, evanescent waves with different decay rates are created inside a diffusive waveguide by the constriction.

In the bow-tie waveguide, the number of transmission eigenchannels that diffuse through the constriction without being converted to evanescent waves is determined by the width of the constriction. When the constriction width $W_c$ is reduced to below the transport mean free path $\ell$, light propagation in the vicinity of the constriction is changed from 2D diffusion to quasi-1D diffusion.  However, the number of waveguide modes supported by the constriction can still be much larger than 1, as long as $W_c \gg \lambda$, allowing light diffusion through the constriction. However, if $W_c < \lambda$, light transport at the constriction changes to evanescent tunneling.

\begin{figure}[htbp]
\centering
\includegraphics[width=3.5in]{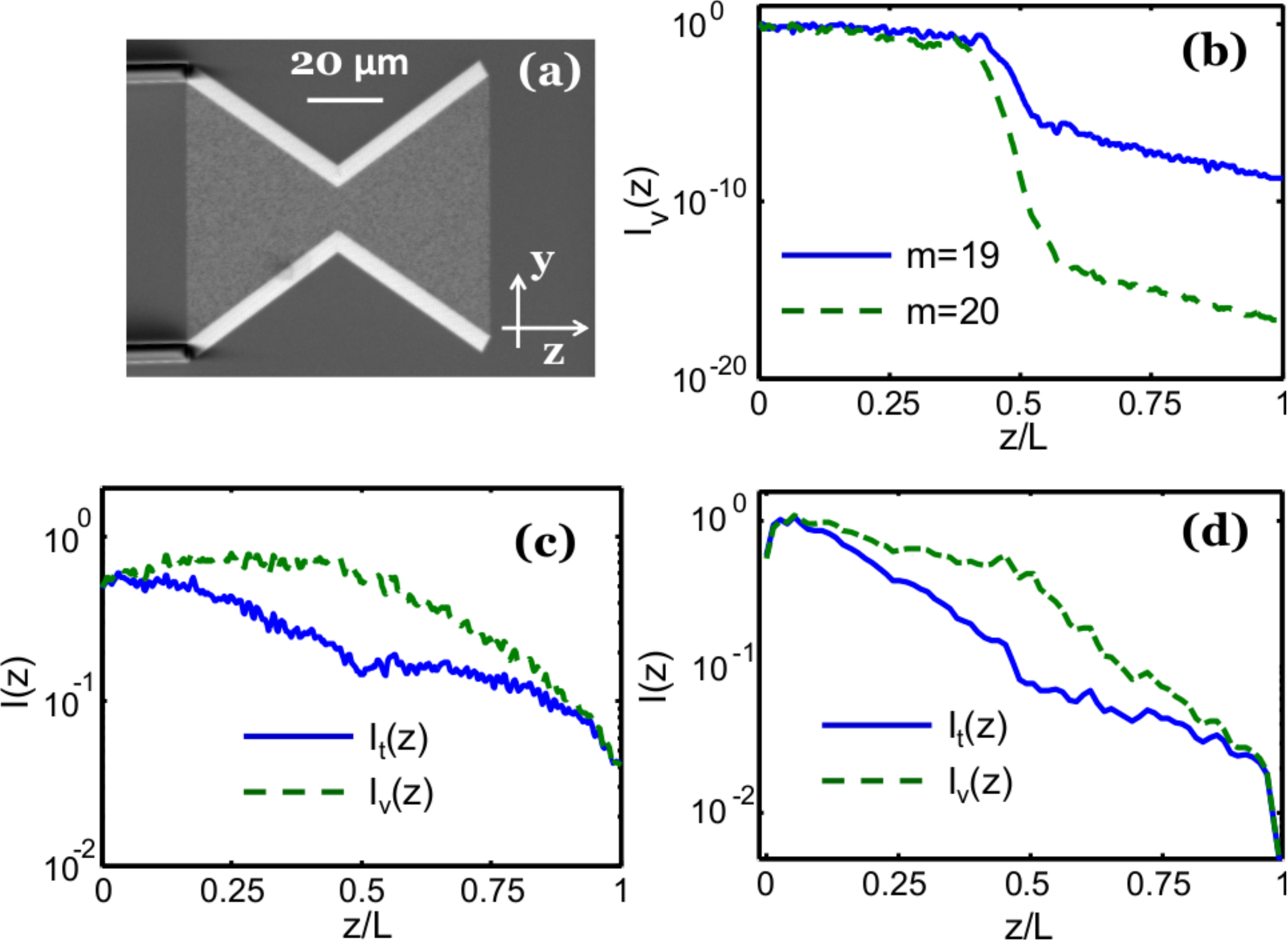}
\caption{(Color online) Transmission eigenchannels and intensity decay in a diffusive waveguide of bowtie geometry. (a) Top-view SEM image of a fabricated quasi-2D waveguide with bowtie geometry. The length of waveguide is $L=80$ $\mu$m. The width of waveguide decreases linearly from 60 $\mu$m at $z=0$ to 10 $\mu$m at $z=L/2$ and then again increases linearly to 60 $\mu$m at $z=L$. (b) Numerically-calculated cross-section-averaged intensity $I_v(z)$ of the 19th (solid line) and 20th (dashed line) transmission eigenchannels of bowtie waveguide. The length $L$ of the waveguide is 20 $\mu$m, the width at $z=0, L$ is 10.2 $\mu$m (35 propagating modes) and the width of constriction at $z=L/2$ is 5.1 $\mu$m (17 propagating modes). The abrupt changes in the decay rate of $I_v(z)$ before and after $z=L/2$ indicate the conversion from propagating wave to evanescent wave and back. The evanescent decay rate varies from one channel to another. (c) Numerically-calculated $I_v(z)$ (green dashed line) and cross-section-integrated intensity $I_t(z)$ (blue solid line) for the maximum transmission channel of the same waveguide as in (b) but with absorption $L/\xi_a = 3$. The constriction causes a significant change in the intensity distribution of the maximum transmission channel. (d) Experimentally measured $I_v(z)$ (green dashed line) and $I_t(z)$ (blue solid line) inside the disordered waveguide shown in (a). Both intensity distributions follow those of the maximum transmission channel.
}
\end{figure}

The bow-tie geometry also modifies the high transmission channels, even in the presence of strong absorption. As shown in Fig. 6(c), the cross-section-averaged intensity $I_v(z)$ for the maximum transmission channel exhibits a small bump at the constriction, while the cross-section-integrated intensity $I_t(z)$ has a dip. This is because the reduction in the cross-section increases the energy density but suppresses the total flux at the center of the waveguide. Figure 6(d) plots the experimentally measured intensity of light inside the bow-tie waveguide. $I_v(z)$ decays slower in the first half ($z< L/2$) than in the second half ($z> L/2$). $I_t(z)$ exhibits opposite behavior. The qualitative agreement between the measured intensity decay and the calculated profile of the highest transmission eigenchannel again confirms that the energy distribution inside the bow-tie waveguide is determined by the maximum transmission channel.

The spatial structure of the open channel in the bow-tie waveguide can be tuned by shifting the constriction away from the center of the waveguide. Unlike varying the constriction width which would modify the transmission eigenvalue and the dimensionless conductance, changing the location of the constriction only modifies the transmission eigenchannels, but not the eigenvalues. It thus provides an efficient way of tailoring the energy distribution inside the diffusive waveguide while keeping the transmittance constant.

Complementary to the bow-tie waveguide, we fabricate the ``lantern'' waveguide whose width $W$ increases linearly in the first half and decreases in the second half [Fig. 7(a)]. In contrast to the bowtie geometry, the number of propagating modes that can be supported in the lantern waveguide increases in the middle due to larger cross-section, thus increasing energy throughput. In particular, a transmission eigenchannel, which is evanescent at the input end of the random waveguide (due to the refractive index difference from the lead waveguide), transforms to propagating wave as the waveguide becomes wider. However, close to the rear end of the waveguide, the propagating wave becomes evanescent again due to the decrease of the waveguide width. Such behavior is shown in Fig. 7(b), where $I_v(z)$ for the $m=19$ eigenchannel exhibits a fast decay near the front end of the lantern waveguide, then the decay is slowed down in the middle, but near the back end the decay becomes fast again. Since light can only tunnel out of the waveguide, there is a strong buildup of energy inside the lantern waveguide, especially near the center where the number of waveguide modes is maximum. For comparison, $I_v(z)$ of another transmission eigenchannel ($m=5$) is also plotted. Unlike $m=19$, $I_v(z)$ for $m=5$ eigenchannel does not display a dip in the intensity at $z/L \sim 0$ as it does not start with an evanescent wave at the front side of the waveguide, instead it exhibits a uniform decay of intensity across the entire waveguide.

The high transmission channels also experience a significant change in the lantern waveguide. As seen in Fig. 7(c), the maximum transmission channel displays an opposite behavior to that of the bow-tie waveguide [Fig. 6(c)]. $I_v(z)$ drops faster in the first half of the waveguide ($z<L/2$) than in the second half ($z> L/2$), while $I_t(z)$ is the opposite. The difference from the bow-tie waveguide is expected because the cross-section is modulated in opposite manner in the two waveguides. Consequently, the intensity distribution inside the lantern waveguide is very different from that in the bow-tie waveguide. The measured $I_v(z)$ and $I_t(z)$ in Fig. 6(d) exhibit distinct decay rates for $z<L/2$ and $z>L/2$, which agree qualitatively to those of the maximum transmission channel. This confirms the change in energy distribution inside the lantern waveguide can be very well represented by the change in the structure of the maximum transmission channel by geometry.

\begin{figure}[htbp]
\centering
\includegraphics[width=3.5in]{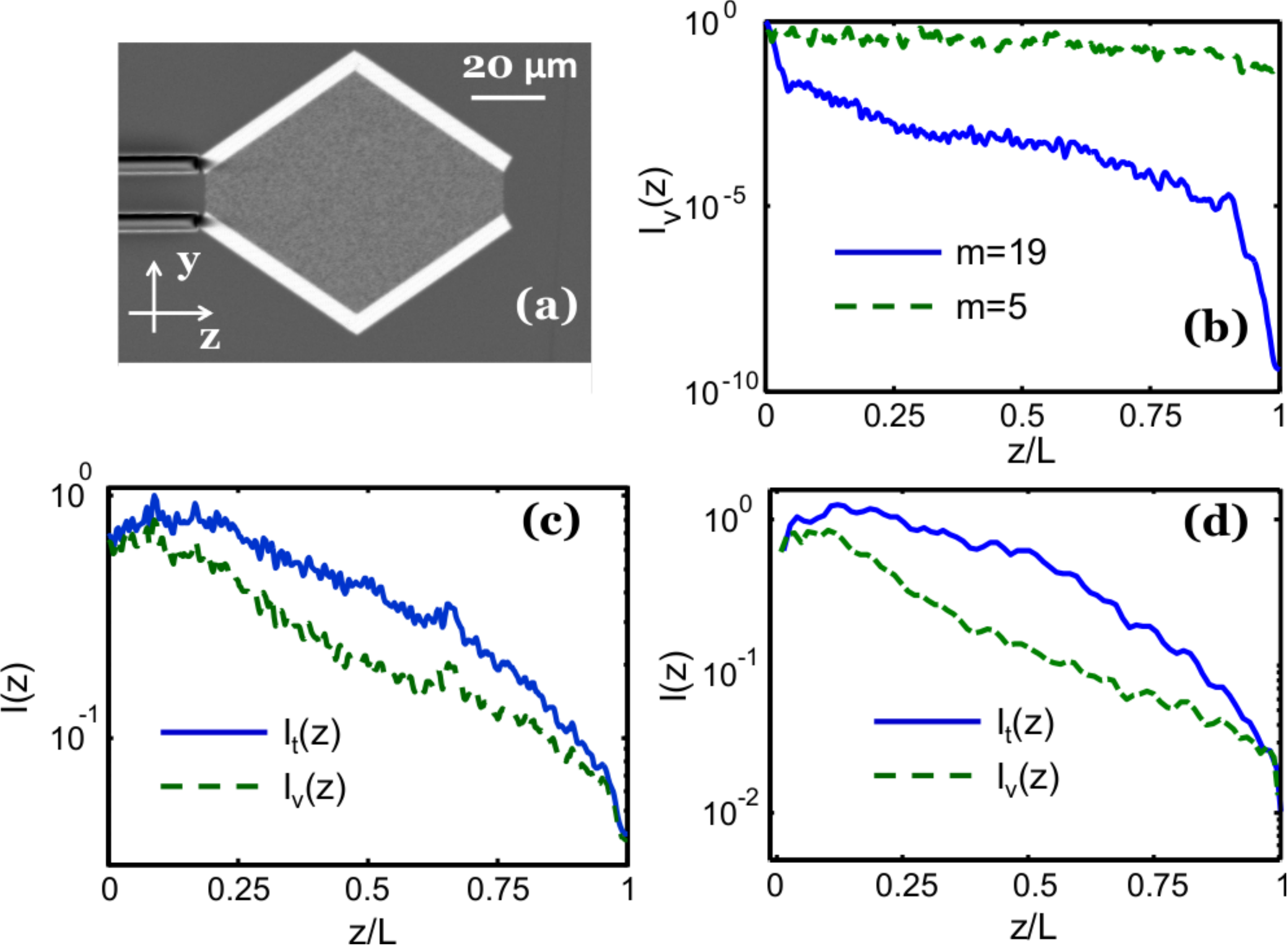}
\caption{(Color online) Transmission eigenchannels and energy distribution in a diffusive waveguide of lantern geometry. (a) Top-view SEM image of a fabricated quasi-2D disordered waveguide with lantern geometry. The length of waveguide is $L=80$ $\mu$m. The width of waveguide increases linearly from 10 $\mu$m at $z=0$ to 60 $\mu$m at $z=L/2$ and then again decreases linearly to 10 $\mu$m at $z=L$. (b) Numerically calculated cross-section-averaged intensity $I_v(z)$ for the 5th (dashed line) and 19th (solid line) transmission eigenchannels of lantern geometry. The length $L$ of the waveguide is 20 $\mu$m, the width at $z=0, L$ is 5.1 $\mu$m (17 propagating modes) and width at $z=L/2$ is 10.2 $\mu$m (35 propagating modes). $I_v(z)$ of the 19th transmission eigenchannel exhibits the conversion of the evanescent wave to a propagating wave near the input end and then back to the evanescent wave near the output end due to the variation of the waveguide width. In contrast, the 5th channel remains propagating wave across the entire waveguide. (c) Numerically calculated $I_v(z)$ (green dashed line) and cross-section-integrated intensity $I_t(z)$ (blue solid line) for the maximum transmission channel of the same waveguide as in (b) but with absorption $L/\xi_a = 3$. Both intensity profiles are opposite to those in the bow-tie waveguide. (d) Experimentally measured $I_v(z)$ (green dashed line) and $I_t(z)$ (blue solid line) inside the disordered waveguide shown in (a). The intensity profiles are similar to those of the maximum transmission channel shown in (c).
}
\end{figure}

\section{Discussion and Conclusions\label{sec:conclusions}}

To conclude, we have demonstrated an effective approach to modify transmission eigenchannels of confined disordered media. Using geometry, we can change the spatial profiles of the transmission channels significantly and deterministically from the universal one. It allows us to control the depth profile of the total energy as well as the energy density inside the random medium, thus controlling how much energy is concentrated inside the sample and where it is concentrated. The ability to tailor the spatial distribution of energy density can be exploited to manipulate light-matter interactions in scattering media, which will be useful for numerous applications.

By gradually increasing the cross-section, we can enhance the transmittance of all the transmission eigenchannels while keeping the number of input modes the same. Such geometries can be useful for applications related to enhancement of total transmission by shaping the input wavefront, as in such structures there will be more open channels due to larger conductance. Moreover, since the waveguide cross-section at the input end does not change, the number of input channels remains the same, and additional degree of control of the input field is not necessary for coupling into any one of the open channels. In addition, using geometry we can also convert evanescent channels to propagating channels and vice versa. In a waveguide with the output cross-section smaller than the input one, perfectly reflecting channels are created. The light injected to such a channel would penetrate inside the scattering media to a certain depth and then get fully reflected back to the input end. The penetration depth of such channels can be further tuned by geometry. Such channels have potential applications for probing deep inside turbid media. Since all the light exits from the input end, the collection efficiency of probe signal would be $100\%$. We can further design geometries with opposite taperings to have the same transmission eigenvalues but very different eigenchannel profiles. By breaking the reflection symmetry of confined geometry, the transmission eigenchannels become asymmetric. In the presence of dissipation, the decay of energy flux inside the diffusive waveguide can be changed by modulating the cross-section of the waveguide along its axis. In a diffusive waveguide with non-monotonic tapering boundary such as the lantern geometry, energy can buildup inside the random medium,  which will benefit the applications of energy harvesting and tailoring of optical excitations inside scattering media.

Unlike the localization effects which are suppressed by absorption, our approach of using geometry to control light transport is effective even in the presence of strong absorption and does not require any change of structural disorder. Thus this approach can truly complement the wavefront shaping technique to control mesoscopic transport of light with an additional advantage that the efficiency is not reduced by external factors such as incomplete channel control and measurement noise \cite{Popoff2014}. The results discussed in this paper are also applicable to other waves such as microwaves, acoustics or matter waves.

Finally, we stress that the confined geometry enables manipulating the spatial structures of transmission eigenchannels while retaining the open channels with perfect transmission. This is advantageous compared to other approaches that rely on strong localization effects, absorption \cite{Liew2014} or asymmetric surface reflections from edges \cite{Tianeigenchannel} to modify the transmission channels as those approaches will also remove the open channels with perfect transmission. Although in this paper we have focused only on the maximum transmission channel, in general using geometry the spatial profiles of the other low transmission channels can also be deterministically and significantly modified. Since changing the confined geometry of a random medium corresponds to modifying its boundary condition, we expect that the Green's function inside the random system can also be tailored. This implies that our approach of manipulating geometry may be applied to control various mesoscopic effects that depend on the Green's functions inside the random media such as non-local intensity correlations, renormalization of the diffusion coefficient, the density of states etc.

\section{Acknowledgments}
We thank Michael Rooks for suggestions regarding sample fabrication. This work is supported by the National Science Foundation under Grants No. DMR-1205307 and DMR-1205223. Facilities use is supported by YINQE and NSF MRSEC Grant No. DMR-1119826.


%

\end{document}